\documentclass[9pt,shortpaper,twoside,web]{ieeecolor}
\pdfoutput=1
\usepackage{generic}
\usepackage{cite}
\usepackage{amsmath,amssymb,amsfonts}
\usepackage{algorithmic}
\usepackage{graphicx}
\usepackage{textcomp,nicefrac}
\usepackage[switch]{lineno}
\usepackage{booktabs}
\usepackage{float}
\usepackage{multirow}
\usepackage{etoolbox}
\usepackage{caption}
\usepackage{amsmath}
\UseRawInputEncoding
\makeatletter
\patchcmd{\@makecaption}
  {\scshape}
  {}
  {}
  {}
\makeatletter
\patchcmd{\@makecaption}
  {\\}
  {.\ }
  {}
  {}
\makeatother

\def\BibTeX{{\rm B\kern-.05em{\sc i\kern-.025em b}\kern-.08em
    T\kern-.1667em\lower.7ex\hbox{E}\kern-.125emX}}    
\begin{document}
\title{Rise Time and Charge Collection  Efficiency of Graphene-Optimized 4H-SiC PIN Detector}
\author{Zhenyu. Jiang, Xuemei. Lu, Congcong. Wang, Yingjie. Huang, Xiaoshen. Kang, Suyu. Xiao, Xiyuan.~Zhang, Xin.~Shi and Hui.~Liang
\thanks{This work is supported by the National Natural Science Foundation of China (Nos. 12305207, 12375184, 12205321 and 12405219), National Key Research and Development Program of China under
Grant No. 2023YFA1605902 from the Ministry of Science and Technology, Natural Science Foundation of Shandong Province Youth Fund
(ZR2022QA098), Liaoning Provincial Department of Education Efficient Basic Research Project (LJ212410140037) and support from CERN DRD3 Collaboration. (Corresponding author: Congcong Wang)}
\thanks{Zhenyu Jiang is with the Institute of High Energy Physics, Chinese Academy of Sciences, Beijing
100049, China,  and also with Liaoning University, Liaoning 110136, China.}
\thanks{Xuemei Lu is with Liaoning University, Liaoning 110136, China.}
\thanks{Zhenyu Jiang and Xuemei Lu contributed equally to this work.}
\thanks{Congcong Wang is with
the Institute of High Energy Physics, Chinese Academy of Sciences, Beijing
100049, China. (e-mail: wangcc@ihep.ac.cn).}
\thanks{Yingjie Huang is with Jilin University, Changchun, 130012, China}
\thanks{Xiaoshen Kang is with Liaoning University, Liaoning 110136, China.}
\thanks{Suyu Xiao and Hui Liang are with Shandong Institute of Advanced Technology, Jinan 250100,China.}
\thanks{Xiyuan Zhang and Xin Shi are with the Institute of High Energy Physics, Chinese Academy of Sciences, Beijing 100049, China.}}
\maketitle

\begin{abstract}
Silicon carbide detectors exhibit good detection performance and have been studied for various detection applications.
However, in some applications the presence of metal is undesirable, such as low-penetration particle detection, UV light detection and medical
dosimetry. A graphene-optimized 4H-SiC detector has been fabricated not only meet the aforementioned detection requirements, but also shorten the signal rise time. Its electrical properties, rise time and the charge collection performance of $\alpha$ particles are reported. The effective doping concentration of lightly doped 4H-SiC epitaxial layer is about 4.5 $\times$ 10$^{13}$cm$^{-3}$, approaching the limit of the lowest doping level by the SiC epitaxial growth technique. The rise time of the graphene-optimized ring electrode detector is reduced by 24\% at 200~V, compared to ring electrode detector. The charge collection efficiency (CCE) of graphene-optimized 4H-SiC PIN is 99.22\%. 
When the graphene has been irradiated using 80 MeV proton
beam with fluence of 2.1~$\times$~10$^{11}$ $\text{n}_{\text{eq}}/\text{cm}^{2}$, the irradiation has no significant impact on the rise time and uniformity of the rise time for the graphene-optimized 4H-SiC detectors. This study proves that graphene has a certain radiation resistance. Graphene-optimized 4H-SiC detectors can not only reduce the signal rise time, but also improve uniformity of signal rise time and stability of charge collection. This research may expand the application of graphene-based 4H-SiC detectors in fields such as low-penetration particles detection, high-energy particles detection, low-energy heavy-ion detection, medical dosimetry and transient current technique (TCT) measurement.
\end{abstract}


\section{Introduction}
\IEEEPARstart Silicon carbide (SiC) detectors have the advantages of higher radiation tolerances\cite{1,2}, lower temperature sensitivity\cite{3}, higher breakdown voltage, low leakage current\cite{4}, efficient charge collection\cite{5} and fast time resolution\cite{4,6} compared with silicon detector.  Silicon carbide detectors exhibit good detection performance and have been studied for various detection applications\cite{4}. The traditional SiC radiation detectors use metal structure electrodes to collect the signals generated by ionising impinging radiation. However,
in some applications the presence of metal is undesirable, such as low-penetration particle detection (low-energy $\alpha$ particles), low-energy heavy-ion detection\cite{7}\cite{10}, the TCT measurement and medical dosimetry. For instance, $\alpha$ particles passing through a thin gold foil and interacting with it cause energy loss, which in turn leads to the broadening of the energy spectrum\cite{26}. Low-energy heavy-ion will cause energy loss when penetrating metals. The radiation detector needs optical windows (without metal) for illumination with a laser source when characterizes its performance of charge collection, position resolution, and time resolution using the TCT\cite{29}. In medical dosimetry such as radiotherapy dosimeters detectors are mainly used for dose monitoring and image guidance. Metals may cause X-ray scattering, thereby affecting the accuracy of the treatment\cite{7}. To reduce this effect, it is necessary to remove a portion of the metal electrodes on surface of the detector, such as fabricated the ring electrode (RE) detector. However, the uneven electric field distribution of the RE detector affects its charge collection and time resolution performance. Therefore, we need a material to enhance the performance of the RE detector and meet the aforementioned applications.

The two-dimensional graphene has attracted great attention due to the excellent properties: Dirac Fermions, high thermal conductivity (5300 W/m·K) and high carrier mobility (200000 cm$^{2}$/~(V·s))\cite{8}. The graphene lattice irradiated by ion and proton has no obvious structural damage\cite{9}. Graphene insertion layer can reduce the metal/4H-SiC interface barrier\cite{11,12}, improve SiC-metal electrical contact as well as the overall thermal management of the devices\cite{8,10,13,14,15}. Graphene can be used as an electrode in the active area of silicon carbide radiation detector to meet the application requirements of low-penetration particle detection (low-energy $\alpha$ particles), low-energy heavy-ion detection, TCT measurements and medical dosimetry\cite{7}. The most important thing is that graphene can improve the signal response time, charge collection stability and time resolution of radiation detectors\cite{7,13,16}. Fast time resolution detectors have significant applications in the field of high-energy physics. Signal rise time and charge collection performance are important parameters of semiconductor detector, which affect the time resolution.

In this work, the ring electrode (RE) 4H-SiC PIN, the surface electrode (SE) 4H-SiC PIN and the graphene-optimized 4H-SiC PIN detector such as graphene/ring electrode (G/RE) have been fabricated. The G/RE 4H-SiC PIN detector requires not only a uniform electric field to meet the aforementioned detection requirements, but also shorten the signal rise time. The RE 4H-SiC PIN and SE 4H-SiC PIN detectors were fabricated to separately analyze the improvement in the performance of the G/RE 4H-SiC PIN detector and whether its performance is equal to or superior to that of the SE 4H-SiC PIN detector. Their electrical characteristics were systematically evaluated in current-voltage (I-V) and capacitance-voltage (C-V) curves. The effect of graphene layer replacing metal contact in the active region of 4H-SiC radiation detector was investigated. Then, the graphene was irradiated first and transferred to the detector to study the influence of radiation graphene on the signal response time and the stability of charge collection. 

\section{Detector fabrication and irradiation conditions}

\subsection{Epitaxial structure and detector fabrication}

The RE, SE, unirradiated G/RE and irradiated G/RE 4H-SiC PIN detectors were fabricated. The size of the detectors is 2mm $\times$ 2mm. The structure of the G/RE 4H-SiC PIN includes monolayer graphene, P electrode, SiO$_{2}$ passivation layer, P++ layer, N-epi layer, N buffer layer, conductive N-type 4H-SiC substrate and N electrode as shown in Fig.~1 (a). The P++ layer with an aluminum ion doping concentration of 2 $\times$ 10$^{19}$ cm$^{-3}$ and a thickness of 0.6 $\mu$m. The lightly doped N-epi layer with a nitrogen ion doping concentration of 5 $\times$ 10$^{13}$ cm$^{-3}$ and a thickness of 50~$\mu$m. The fabrication process of the G/RE 4H-SiC PIN detector mainly includes lithography, etching, electron beam evaporation, magnetron sputtering, rapid thermal annealing, transfer and etch graphene. Etching depth of epitaxial structure is more than 0.6~$\mu$m to ensure full etching of P++ layer. The Ni/Ti/Al = 50nm/15nm/60nm as the electrode was grown on the top of P++ layer and N-type substrate by using electron evaporating method. The rapid annealing time and temperature are 2min and 950~$^\circ$C to form P-ohmic contact. A SiO$_{2}$ layer with a thickness of 500 nm was deposited by plasma-enhanced chemical vapor deposition (PECVD) at 350~$^\circ$C. The most difficult process is transfering and etching graphene. Firstly, the copper-based graphene (SixCarbon Technology) is irradiated. Graphene irradiation conditions is presented in TABLE~I. The copper-based monolayer graphene is coated with polymethyl methacrylate (PMMA) solution. Then, ferric chloride (FeCl$_{3}$) solution is used to etch away the copper foil, a PMMA film with graphene is transferred to the detector surface. Finally, the PMMA is removed with anisol. After the graphene transfer is completed, the graphene patterning is obtained by photolithography, development and reactive ion etching (RIE).
The purposes of graphene patterning are to remove the graphene in wire-bonding area and break the graphene electrode connections between the detectors.

\begin{figure}[htbp]
\centering
\includegraphics[scale=1]{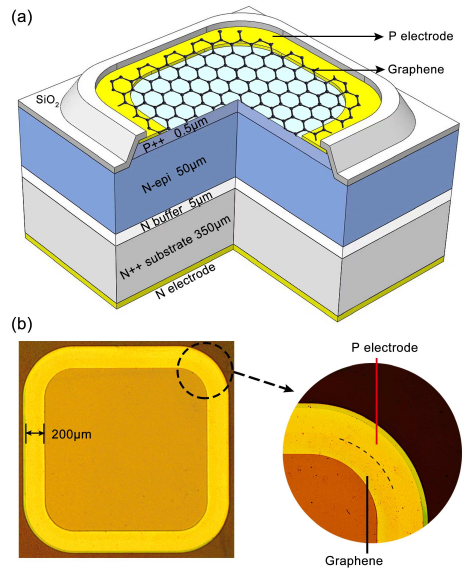}
\caption{(a) 3D cross-sectional schematic of the G/RE 4H-SiC PIN radiation detector. (b) Real G/RE 4H-SiC PIN radiation detector image.}
\label{figure}
\end{figure}

\subsection{Irradiation conditions}
The graphene irradiation was conducted at the Associated Proton Beam Experiment Platform (APEP) beamline of the China Spallation Neutron Source (CSNS) in Dongguan, Guangdong Province, China. We irradiate the graphene at the test point with an energy of 80 MeV and a beam spot size of 20 mm$\times$20 mm. The tested flux of protons at the test point for a beam spot of this size is 5.59~$\times$~10$^{9}$~p/cm$^{2}$/s. The NIEL (non-ionizing energy loss) factor for an 80 MeV proton beam is calculated to be about 1.427 to be equivalent to 1-MeV neutron\cite{30,31}. They  respectively take around 26s and 4430s to reach 1.45$\times$~10$^{11}$~p/cm$^{2}$ and 2.48$\times$~10$^{13}$~p/cm$^{2}$, corresponding to the fluences of 2.1~$\times$~10$^{11}$~$\text{n}_{\text{eq}}/\text{cm}^{2}$ and 3.5~$\times$~10$^{13}$~$\text{n}_{\text{eq}}/\text{cm}^{2}$. The irradiation environment is at room temperature. Detailed information of detectors and graphene irradiation conditions are presented in TABLE I. 

\begin{table}[H]
 \caption{Information of 4H-SiC PIN detectors and graphene irradition fluences}
 \centering
 \resizebox{88mm}{14mm}{
 \begin{tabular}{ccc}
    \toprule    
    \multirow{2}{*}{Label} & \multirow{2}{*}{Electrode} & Graphene irradition fluence \\
    &&($\text{n}_{\text{eq}}/\text{cm}^{2}$) \\
    \midrule    
    RE PIN & Ring electrode& 0 \\
    SE PIN & Surface electrode & 0 \\
    G/RE PIN-1 & Ring electrode+Graphene & 0 \\
    G/RE PIN-2 & Ring electrode+Graphene & 2.1 $\times$~10$^{11}$ \\
    G/RE PIN-3 & Ring electrode+Graphene & 3.5 $\times$~10$^{13}$ \\
    \bottomrule
 \end{tabular}
}
\end{table}

\subsection{Conductive atomic force microscopy measurement}
Graphene/4H-SiC interface was investigated by current transport across the interface. Conductive atomic force microscopy (C-AFM) measurements can be used to verify whether the contact between graphene and silicon carbide has conductive properties. The Oxford Cypher S AFM and conductive diamond probes (AD-2.8-AS) were used in the measurement. Fig. 2~(a) shows a standard measurement setup used for current spectroscopy in an air environment. If graphene has conductive properties when in contact with 4H-SiC, current characteristics can be obtained when a conductive probe contacts graphene and a bias voltage is applied. Fig. 2 (b) shows the I-V curve of graphene at point A in Fig. 2 (c). It can be seen from the I-V curve that as the voltage increases, the current increases. Fig. 2 (c) shows the morphology image of atomic force microscopy (AFM). Fig. 2 (d) shows the current mapping of graphene deposited on 4H-SiC at the same position as in Fig. 2 (c) when the voltage is 2~V. The I-V curve and current map indicate that graphene conducts electricity when in contact with silicon carbide, and graphene can be used as an electrode material.

\begin{figure}[htbp]
\centering
\includegraphics[scale=1]{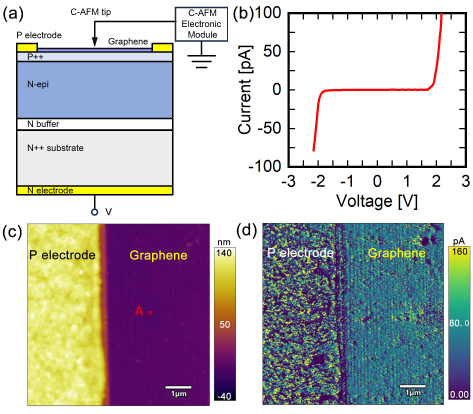}
\caption{(a) The C-AFM measurement setup. Bias is applied at the back end of the detector while a conductive tip carries current to the ground through a C-AFM current sensor. (b) The I-V curve of graphene at point A in Fig. 2 (c). (c) Morphology image of AFM. (d) The current at the same position as in Fig. 2 (c) when the voltage is 2 V.}
\label{figure}
\end{figure}

\section{Electrical performance analysis}

\begin{figure}[htbp]
\centering
\includegraphics[scale=1]{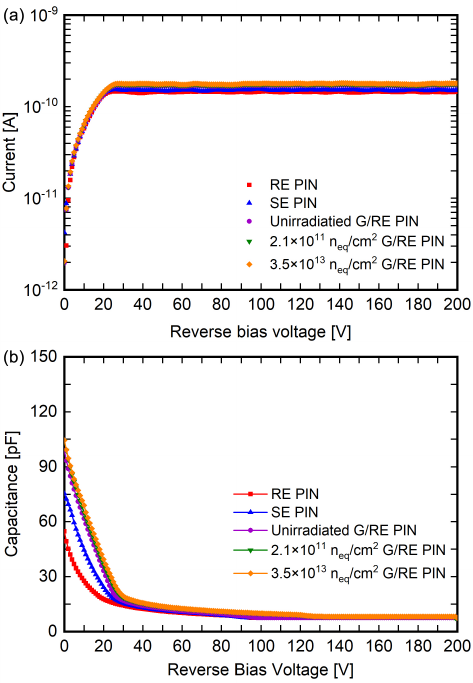}
\caption{The RE 4H-SiC PIN, SE 4H-SiC PIN, unirradiated G/RE 4H-SiC PIN detectors, 2.1 $\times$ 10$^{11}$ $\text{n}_{\text{eq}}/\text{cm}^{2}$ and 3.5 $\times$ 10$^{13}$ $\text{n}_{\text{eq}}/\text{cm}^{2}$ irradiated G/RE 4H-SiC PIN detectors: (a) I-V characteristics. (b) C-V characteristics.}
\label{figure}
\end{figure}

The I-V characteristics were measured on a probe station at room temperature, using a Keithly 2470 source meter. The C-V characteristics were measured under same environment with Keysight E4980 LCR meter and a bias adapter, at frequency f=10kHz. The I-V characteristics of the RE, SE, unirradiated G/RE and irradiated G/RE 4H-SiC PIN detectors are shown in Fig. 3 (a). The leakage currents and leakage current densities of the RE, SE and G/RE 4H-SiC PIN are about 0.147~nA (3.675~nA/cm$^{2}$)~@~200~V, 0.148~nA (3.700~nA/cm$^{2}$)~@~200~V and 0.171~nA (4.275~nA/cm$^{2}$)~@~200~V, respectively. The surface of the G/RE 4H-SiC PIN is covered with graphene without a passivation layer, so the leakage current and leakage current density are higher than the other two devices. The C-V characteristics of the RE, SE, unirradiated G/RE and irradiated G/RE 4H-SiC PIN detectors are shown in Fig. 3 (b). The C-V curves show that the full depletion voltage of these detectors is about 120~V. The effective doping concentration and depletion depth of N-epi layer are about 4.5 $\times$ 10$^{13}$~cm$^{-3}$ and 45~$\mu$m, respectively. The effective doping concentration approaching the lowest doping level in SiC epitaxial growth technique\cite{17,18}. The doping concentration and thickness meet the design requirements. It can be seen from the Fig.~3, whether the graphene is irradiated or not basically does not affect the electrical performance of the detectors.

\section{Rise time and charge collection performance analysis}

\subsection{Experimental setup}

The charge collection performance setup for $\alpha$ particles is shown in Fig. 4. The system includes $^{241}$Am radioactive source, 4H-SiC PIN, single channel electronic readout board, high voltage source (Keithley 2470), low voltage source (GPD-3303, SGWINSTE) and oscilloscope (MSO64, Tektronix 2.5 GHz). The 4H-SiC detectors are encapsulated on the electronic readout board by using conductive adhesive, and the pad electrode of 4H-SiC detectors are connected to the readout board. The high voltage source provides the detector with reverse bias. The low voltage source provides power supply to the single channel electronic readout board. The current signal is generated when the particles emitted by the radioactive source pass through the device. The current signal is amplified by an electronic amplifier and converted into a voltage signal. Then the voltage signal is collected by oscilloscope to form pulse waveform. The rise time distribution of the waveform can be obtained. Finally, the charge collection information is obtained by integrating the pulse waveform and converting the current-time signal.

\begin{figure}[htbp]
\centering
\includegraphics[scale=1]{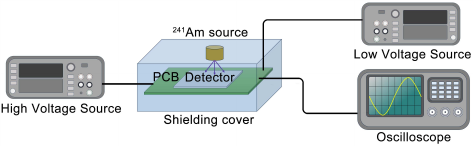}
\caption{Experimental setup for $\alpha$ particle measurement}
\label{figure}
\end{figure}

\subsection{Rise time and charge collection performance of the RE, SE and G/RE 4H-SiC PIN}

 Signal rise time and charge collection performance are important parameters of semiconductor detector, which affect the time resolution and efficiency of detector. Although the C-V curves show that the full depletion voltages are about 120~V. To shorten the signal rise time, we study the performances of the detectors at 200~V. Fig.~5~(a) (b) and (c) show the signal waveforms and rise time distributions of the RE, SE and unirradiated G/RE 4H-SiC PIN detectors at 200~V. The rise time and sigma for the RE, SE and unirradiated G/RE 4H-SiC PIN are 443~ps, 335~ps, 336~ps and 84.7, 31.4, 41.0, shown in the Fig. 5~(a) (b) and (c). The rise time of the G/RE 4H-SiC PIN detector is reduced by 24\% at 200~V, compared to RE 4H-SiC PIN detector. The experiments prove that compared with the RE 4H-SiC PIN detector, the graphene-optimized RE 4H-SiC PIN detector, namely the unirradiated G/RE 4H-SiC PIN,  has a fast and uniform rise time. The electric field distribution of the SE PIN detector is uniform. The rise time and uniformity of the rise time for the SE 4H-SiC PIN detector are better. However, the overlap phenomenon that may exist due to the inconsistent height between graphene and the metal electrode makes the charge collection uniformity of G/RE 4H-SiC PIN slightly worse than that of SE 4H-SiC PIN detector.

\begin{figure}[H]
\centering
\includegraphics[scale=1]{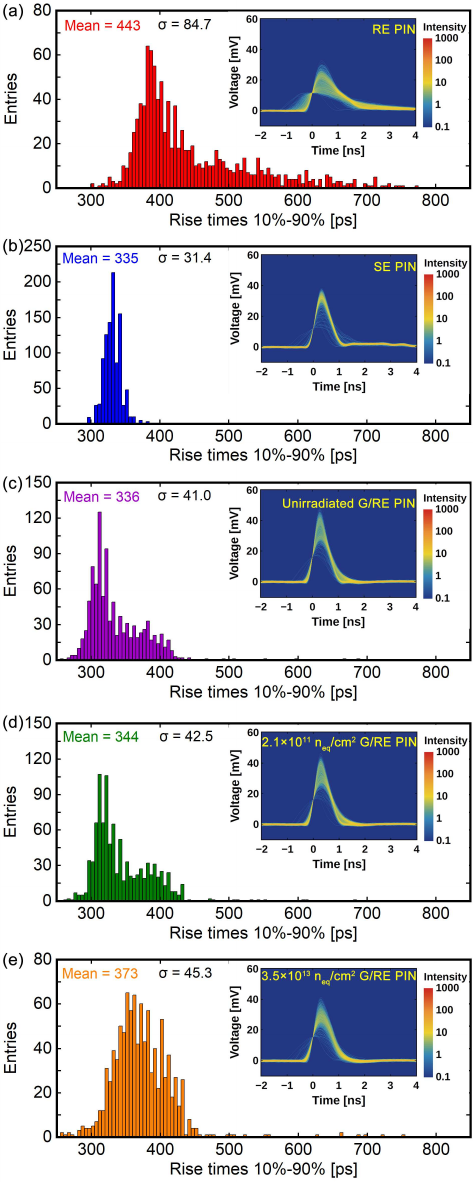}
\caption{Signal waveforms and rise time distributions at 200~V: (a) RE 4H-SiC PIN. (b) SE 4H-SiC PIN. (c) Unirradiated G/RE 4H-SiC PIN. (d) 2.1~$\times$~10$^{11}$~$\text{n}_{\text{eq}}/\text{cm}^{2}$ irradiated G/RE 4H-SiC PIN. (e)~3.5~$\times$~10$^{13}$~$\text{n}_{\text{eq}}/\text{cm}^{2}$ irradiated G/RE 4H-SiC PIN.}
\label{figure}
\end{figure}

 As shown in Fig. 6 (a), when particles are incident on the detector, holes are generated and move vertically to the electrode surface for collection. The electric field distribution of the RE PIN detector is uneven, the middle electric field intensity is low, and the electric field intensity is high near the position of electrode ring. As shown in Fig.~6 (b), the holes generated by the particles incident on the center of the detector will move laterally in the P++ layer to the P-type electrode and be collected. The holes moved laterally for a longer distance compared to the vertical movement. The rise time of the RE PIN detector is relatively long and unstable, the signal shows a long tailing and the charge collection performance is unstable. The rise time of the SE PIN detector is short and stable, the charge collection performance is stable. In the G/RE 4H-SiC PIN detector, the graphene electrode in the middle serves to homogenize the electric field. So its rise time is shorter and charge collection performance is relatively stable compared to the RE PIN detector. Therefore, it can be seen that graphene can be used as an electrode material to make the charge collection rise time faster and improve the uniformity of the rise time.

\begin{figure}[htbp]
\centering
\includegraphics[scale=1]{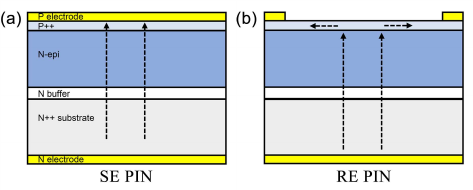}
\caption{Hole motion trajectories: (a) SE 4H-SiC PIN. (b) RE 4H-SiC PIN.}
\label{figure}
\end{figure}

The $^{241}$Am source decays to release $\alpha$ particles with an energy of 5.486 MeV. It can be fully absorbed by detectors with 50 $\mu$m epitaxial layer. The $\alpha$ particles can produce electron-hole pairs after it passes through the depleted layers. The full depletion voltages are about 120~V. A Gaussian distribution function is used to analyze the charge collection for 4H-SiC PIN detectors. For voltages greater than 120~V no further increase of the depletion depth occurs, and no increase in the collected charge is observed. The charge collection distributions of the RE, SE and unirradiated G/RE 4H-SiC PIN detectors at a bias of 200~V is shown as Fig. 7. The collected charges of the RE, SE and unirradiated G/RE 4H-SiC PIN detectors are 58.27~fC, 59.46~fC and 59.00~fC at 200~V.  We define the CCE for the SE PIN detector as 100\%~@~200~V. The CCEs of the RE 4H-SiC PIN and unirradiated G/RE 4H-SiC PIN are 97.99\% and 99.24\% at 200~V, shown in Fig.~7~(b). 

\begin{figure}[htbp]
\centering
\includegraphics[scale=1]{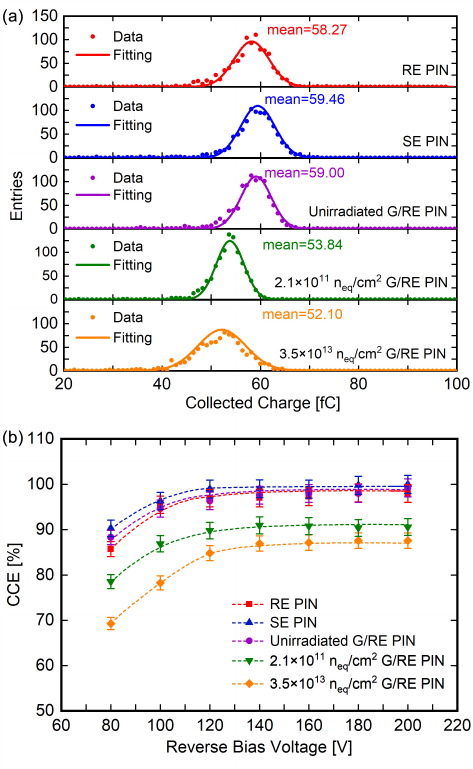}
\caption{The RE, SE, unirradiated G/RE 4H-SiC PIN detectors, 2.1~$\times$ 10$^{11}$ $\text{n}_{\text{eq}}/\text{cm}^{2}$ and 3.5 $\times$ 10$^{13}$~$\text{n}_{\text{eq}}/\text{cm}^{2}$ irradiated G/RE 4H-SiC PIN detectors G/RE PIN detectors: (a) Charge collection distributions at 200V. (b) Charge collection efficiencies.}
\label{figure}
\end{figure}

\subsection{Rise time and charge collection performance of the irradiated G/RE 4H-SiC PIN}

In order to study whether graphene can be applied to the field of radiation-resistant particle detectors. We first irradiate graphene and then transfer it to the detector study the effect of irradiated graphene on the charge collection performance. Fig 5~(d) and (e) show the signal waveforms and rise time distributions of the irradiated G/RE 4H-SiC PIN detectors at 200V. The mean values of rise time and sigma for the 2.1 $\times$ 10$^{11}$ $\text{n}_{\text{eq}}/\text{cm}^{2}$ and 3.5 $\times$ 10$^{13}$ $\text{n}_{\text{eq}}/\text{cm}^{2}$ irradiated G/RE 4H-SiC PIN detectors are 344.0~ps, 372.5~ps and 42.5, 45.3, respectively. The rise times of 2.1 $\times$ 10$^{11}$ $\text{n}_{\text{eq}}/\text{cm}^{2}$ and 3.5 $\times$ 10$^{13}$ $\text{n}_{\text{eq}}/\text{cm}^{2}$ irradiated G/RE 4H-SiC PIN are decreased by 22.3\% and 15.9\% at 200V, compared to the RE 4H-SiC PIN. When the irradiation dose is 2.1~$\times$ 10$^{11}$ $\text{n}_{\text{eq}}/\text{cm}^{2}$, the irradiation has no significant impact on the rise time and uniformity of the rise time for the G/RE 4H-SiC PIN detector. When the irradiation dose is 3.5 $\times$ 10$^{13}$ $\text{n}_{\text{eq}}/\text{cm}^{2}$, the rise time becomes longer and the uniformity of the rise time deteriorates. Fig. 7 shows the charge collection distributions and CCEs of the irradiated G/RE 4H-SiC PIN. The CCEs of the 2.1~$\times$ 10$^{11}$ $\text{n}_{\text{eq}}/\text{cm}^{2}$ and 3.5 $\times$ 10$^{13}$ $\text{n}_{\text{eq}}/\text{cm}^{2}$ irradiated G/RE 4H-SiC PIN are 90.59\% and 87.58\% at 200~V. With increasing irradiation dose, the CCEs are slightly reduced.

\begin{figure}[htbp]
\centering
\includegraphics[scale=1]{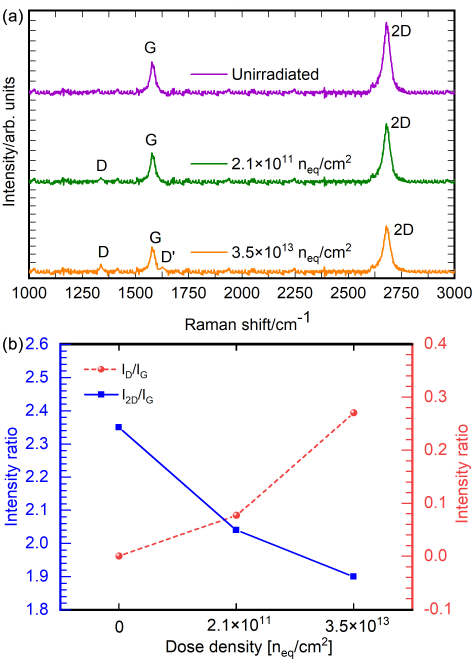}
\caption{ (a) Normalized Raman spectra of graphene. (b) The relationship between Raman peak strength
I$_{D}$/I$_{G}$, I$_{2D}$/I$_{G}$ ratio and irradiation dose.}
\label{figure}
\end{figure}

We can explain the above phenomenon through the irradiation defects of graphene. Raman spectra can analyze the defects of graphene. In this experiment, Raman spectra of graphene were measured at room temperature using LabRam HR80 laser confocal spectrometer. The wavelength, spot radius and power of the laser are 532~nm, 1~$\mu$m and 5~mW, respectively. Fig.8 (a) shows representative Raman spectra of graphene transferred to the 4H-SiC detector. Fig.8 (b) shows the relationship between I$_{D}$/I$_{G}$, I$_{2D}$/I$_{G}$ and irradiation dose. The G peak (1583 cm$^{-1}$) is usually generated from the in-plane vibration of the sp$^{2}$ carbon atoms which originating from the E$_{2g}$ symmetry of in-plane phonons at the $\Gamma$ point of the Brillouin zone. The 2D peak (2671 cm$^{-1}$) is a two-phonon resonant Raman peak related to the interlayer stacking mode of the carbon atoms\cite{19,20}. When the I$_{2D}$/I$_{G}$ ratio is greater than 1.5, which proves that the graphene on 4H-SiC PIN is a single layer graphene\cite{21}. The Peak D (1340 cm$^{-1}$) shows the breathing pattern caused by the defect, which represents the sp$^{3}$ hybrid structure in the carbon atom structure network. The I$_{D}$/I$_{G}$ ratio is commonly used to characterize the important parameter of the defect density of graphene. When the I$_{D}$/I$_{G}$ ratio is larger, it indicates that there are more defects in graphene. As shown in the Fig.8, the graphene has distinct G and 2D peaks. The 2D peaks have perfect single Lorentz peaks, and the I$_{2D}$/I$_{G}$ ratio is greater than 1.5, which proves that the graphene on 4H-SiC PIN is a single layer graphene\cite{21}. When the irradiation dose was 2.1~$\times$ 10$^{11}$~$\text{n}_{\text{eq}}/\text{cm}^{2}$, a weak D peak appeared at 1340 cm$^{-1}$, and the I$_{D}$/I$_{G}$ ratio was 0.077, indicating that there are defects in the irradiated graphene. When the irradiation dose increased to 3.5~$\times$~10$^{13}$~$\text{n}_{\text{eq}}/\text{cm}^{2}$, obvious D and weak D' peaks  \cite{22} appeared at 1340 cm$^{-1}$ and 1620 cm$^{-1}$, respectively. The intensity of D peak increased significantly, the I$_{D}$/I$_{G}$ ratio was 0.27. At this irradiation dose, the defect density and disorder degree of graphene increase, presenting a nanocrystalline structure\cite{20}. In conclusion, with the increase of irradiation dose, the defects of graphene increase.

High energy proton irradiation can produce point defect damage in the graphene layer. The collision of incident protons with lattice atoms can cause displacement damage, possibly result in the deep states within the bandgap\cite{23}. Proton irradiation causes graphene to form charge traps, which reduces the carrier mobility and carrier lifetime of graphene\cite{24,25}. Therefore, the conductivity of graphene decreases and the device electric field changes, which may reduce the detector charge collection efficiency. At the same time, the electrons and holes generated by the energy deposition of $\alpha$ particles in the depletion region may be captured by the defects of graphene, so that the effective electrons and holes cannot be completely collected, and then the CCEs are reduced. Therefore, the G/RE 4H-SiC PIN detector with an irradiation dose of 3.5~$\times$~10$^{13}$ $\text{n}_{\text{eq}}/\text{cm}^{2}$ has the longest and least uniform signal rise time and the lowest CCE, compared to the G/RE 4H-SiC PIN detectors with an irradiation dose of 0 and 2.1~$\times$~10$^{11}$~$\text{n}_{\text{eq}}/\text{cm}^{2}$. An increase in graphene defects may reduce its electrical conductivity and alter the internal electric field distribution of the detector. This may increase the signal rise time of the detector, reduce the uniformity of the signal rise time and the CCE.

In conclusion, graphene can be used as an electrode material to make the rising time faster and improve the uniformity of the rise time. When the irradiation dose is 2.1 $\times$ 10$^{11}$ $\text{n}_{\text{eq}}/\text{cm}^{2}$, the irradiation has no significant impact on the rise time and uniformity of the rise time for the G/RE 4H-SiC PIN detector. With the increase of irradiation dose, the defects of graphene will increase. This leads to an increase in the rise time, a deterioration in the uniformity of the rise time and a decrease in the charge collection efficiency for the detector.

\section{Conclusion}

A graphene-optimized 4H-SiC detector has been fabricated to detect $\alpha$ particles. In this graphene-optimized 4H-SiC PIN detector, the effective doping concentration of lightly doped 4H-SiC epitaxial layer is about 4.5 $\times$ 10$^{13}$ cm$^{-3}$, approaching the limit of the lowest doping level by the SiC epitaxial growth technique. The rise times of unirradiated and 2.1 $\times$ 10$^{11}$ $\text{n}_{\text{eq}}/\text{cm}^{2}$ irradiated graphene-optimized ring electrode 4H-SiC PIN are decreased by 24\% and 22.3\% at 200~V, compared to the RE 4H-SiC PIN. When the irradiation dose is 2.1 $\times$ 10$^{11}$ $\text{n}_{\text{eq}}/\text{cm}^{2}$, the irradiation has no significant impact on the rise time and uniformity of the rise time for the G/RE 4H-SiC PIN detector. This work provides graphene has a certain radiation resistance and a novel structure for application in the field of particle physics. Graphene-optimized 4H-SiC detectors can not only reduce the signal rise time, but also improve uniformity of signal rise time and stability of charge collection.

The graphene-optimized silicon carbide detector has surface ultraviolet laser transmissibility. The TCT technology will be utilized to test the charge collection performance and time resolution performance of this detector. In the future, the graphene-optimized silicon carbide strip and pixel detectors will be fabricated. Their position resolution and time resolution performance will be characterized by TCT. Meanwhile, it opens a possibility to use the graphene-based 4H-SiC radiation detector for application in the field of low-penetration particles detection, high-energy particles detection, low-energy heavy-ion detection, medical dosimetry and TCT measurement.

\section{Acknowledgement}
We acknowledge the RASER team (https://raser.team) and CERN DRD3 Collaboration (https://drd3.web.cern.ch) for their useful discussions.

\bibliographystyle{ieeetr}
\bibliography{sample}

\end{document}